\documentclass[aps,prl,twocolumn,showpacs,amsmath,prepintnumbers,amssymb,superscriptaddress,reprint]{revtex4}
\usepackage{graphicx}
\usepackage{dcolumn}
\usepackage{color}
\usepackage[latin1]{inputenc}
\usepackage{comment}
\usepackage{amsmath}

\usepackage[normalem]{ulem}

\graphicspath{{./Fig1/},{./Fig2/},{./Fig3/}}
\begin{document}
\preprint{Version v8}
%
\title{Generation of Transient Photocurrents in the Topological Surface State
of Sb$_{2}$Te$_{3}$ by Direct Optical Excitation with Mid-Infrared Pulses}
\author{K.~Kuroda}
\author{J.~Reimann}
\author{J.~G{\"u}dde}
\author{U.~H{\"o}fer}
\affiliation{Fachbereich
Physik und Zentrum f{\"u}r Materialwissenschaften,
Philipps-Universit{\"a}t, 35032 Marburg, Germany}
\date{\today}
 \pacs{73.20.-r, 78.47.J-, 79.60.-i, 79.60.Bm}

%
\begin{abstract}                                                   %
%
We combine tunable mid-infrared (MIR) pump pulses with time- and
angle-resolved two-photon photoemission to study ultrafast
photoexcitation of the topological surface state (TSS) of
Sb$_{2}$Te$_{3}$.
 It is revealed that MIR pulses permit a direct excitation of
the unoccupied TSS owing to an optical coupling across the Dirac
point.
 The novel optical coupling provokes asymmetric transient populations
of the TSS at ${\pm}k_{||}$, which mirrors a macroscopic photoexcited
electric surface current.
 By observing the decay of the asymmetric population, we
directly investigate the dynamics of the long-lived
photocurrent in the time domain.
 Our discovery promises important advantages of photoexcitation by MIR pulses for
spintronic applications.
\end{abstract}
\maketitle
%
%
%
Three-dimensional (3D) topological insulators (TIs) belong to a new class of
materials which are characterized by an insulating bulk and a metallic
topological surface state (TSS)~\cite{Hasan10rmp}.
 The most remarkable properties of the TSS are its Dirac-cone-like energy
dispersion and its chiral spin texture in $k$-space~\cite{Xia09Nature,
  Chen09Science}.
 The latter incorporates a protection against backscattering and is
 very promising for spintronic applications.
 Optical coupling to the chiral spin texture of TSSs offers
many interesting phenomena, such as optical control of
the spin~\cite{Joswiak13NaturePhys}, surface
transport~\cite{McIver12NatureNano, Olbrich14PRL, Kastl15NatureCom},
and topological phases~\cite{Wang13Science}.
 To exploit these exotic properties, a detailed understanding
of the optical excitation and the subsequent electron dynamics are essential keys.
 A number of studies have investigated the ultrafast electron dynamics
by optical methods like reflectivity~\cite{Qi2010APL}, second harmonic
generation~\cite{Hsieh2011PRL}, or by optically triggered detection of
photocurrents \cite{Kastl15NatureCom}.
 These experiments, however, hardly show a pure photoexcitation of
TSSs, since the bulk response typically governs the total signal.

Time- and angle-resolved two-photon photoemission (2PPE) is a
particularly suited technique for this purpose~\cite{Bovensiep10},
because it can directly image the optically excited electron
population by energy-momentum mapping and makes it possible to
follow the ultrafast carrier dynamics in the time domain by pump-probe
schemes.
 This technique has been successfully applied for the study of the
electron dynamics in 3D TIs~\cite{Sobota12prl,Wang12prl,Crepaldi12prl,Hajlaoui12Nano,Sobota13prl,
  Niesner14prb,Reimann14prb, Zhu15SientificReport} using pump laser pulses at 800-nm or
in the visible range.
 In contrast to higher-lying states, such as image-potential states,
however, the relevant TSS close to the Fermi energy ($E_{\rm F}$)
is in these studies only indirectly excited by a delayed filling
from states far above $E_{\rm F}$.
 Under such conditions, a coherent optical control of the TSS population
is difficult and its decay dynamics is masked.

In this Letter, we demonstrate that tunable low-energy pump
pulses in the mid-infrared (MIR) regime are capable to induce a
direct optical transition between the occupied and unoccupied
part of the TSS across the Dirac point (DP). We show this for
the example of $p$-doped Sb$_{2}$Te$_{3}$, in which the most
part of its Dirac cone is initially
unoccupied~\cite{Reimann14prb,Zhu15SientificReport}.
 In contrast to a bulk mediated indirect population, this
resonant TSS-TSS transition makes it possible to generate an
asymmetry between the transient population of opposite parallel
momenta $k_{||}$, which directly mirrors a macroscopic spin
polarized photocurrent within the TSS.
 By monitoring the decay of this asymmetry, we identify different scattering
mechanisms of the electrons that carry the photocurrent.

\begin{figure*}[t!]
\begin{center}
\includegraphics[width=0.9\textwidth]{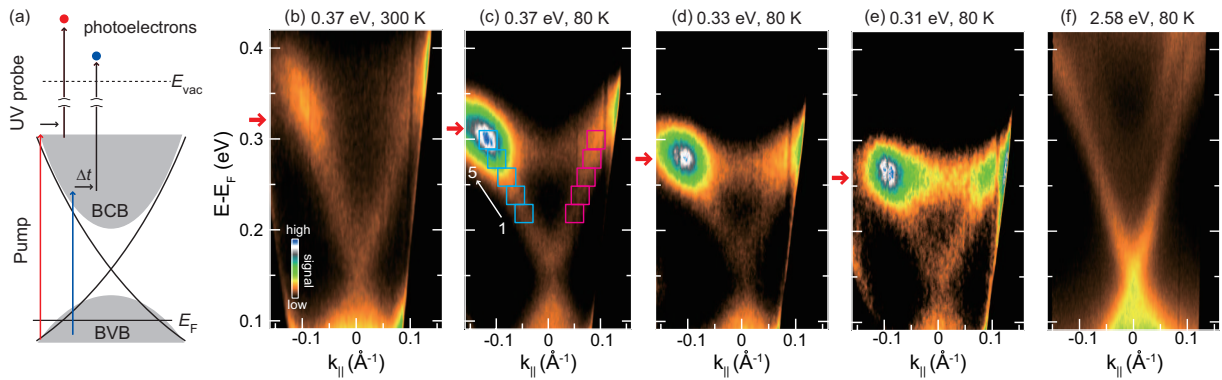}
\caption[]{(Color online)
  (a) Excitation scheme for the population of
  the TSS in Sb$_2$Te$_3$ with MIR pump pulses and subsequent photoemission with
  ultraviolet probe pulses.
  (b) and (c) Angle-resolved 2PPE spectra for
  0.37~eV pump and 5.16~eV probe pulses 50~fs after MIR excitation at 300~K and
  80~K, respectively.  Red arrows indicate the energies
  of maximum population enhancement in the TSS.
  (d) and (e) show the results acquired at 80~K using
0.33~eV and 0.31~eV pump pulses, respectively. (f) shows a spectrum for
  2.58~eV pump pulses for comparison.}
\label{fig1}
\end{center}
\end{figure*}
%
%
%
Details of our optical setup are described in the Supplemental
Material~\cite{SI}.
 Electrons were excited into initially unoccupied states above
$E_{\rm{F}}$ with 100-fs MIR laser pulses of tunable
photon energy ($h\nu_1$=0.25-0.37~eV).
 The transient population was subsequently probed by photoemission of
these electrons using ultraviolet (UV) laser pulses ($h\nu_2$=5.16~eV,
80~fs)~[Fig.~\ref{fig1}~(a)].
 Both $p$-polarized beams were focused on the sample into a spot
with a diameter of $\sim$100~$\mu$m using a non-collinear geometry.
 The experiments were carried out in a $\mu$-metal shielded UHV chamber
at a base pressure of $4\times 10^{-11}$~mbar.
 Photoelectrons were collected along the high symmetry
line $\bar{\rm{\Gamma}}$-$\bar{{\rm{K}}}$ by a hemispherical analyzer
(Specs Phoibos 150) with a display-type detector.
 A single crystal of $p$-doped Sb$_{2}$Te$_{3}$ was cleaved {\it in
  situ} by the Scotch tape method at a pressure of $3\times
10^{-10}$~mbar followed by a rapid recovery back to the base pressure
within a minute.
 The Dirac point (DP) of the sample was located $\sim$150~meV
above $E_{\rm{F}}$.
 During the measurements the sample temperature was maintained at 300~K
or 80~K.

%
%
%
We start by discussing the optical excitation process into the unoccupied TSS
by means of Figures~\ref{fig1}~(b) and (c), which show angle-resolved 2PPE spectra
of Sb$_{2}$Te$_{3}$ using a pump photon energy $h\nu_1$ of
0.37~eV at 300~K and 80~K, respectively.
 To avoid interfering 2PPE signals from the image-potential states,
which are excited by the UV pulses and probed by the MIR pulses,
a 50~fs delay of the UV probe pulses with respect to the MIR pump
pulses has been used.
 Even at this small delay, a considerable population of the TSS can
already be observed.
 It extends only up to 350~meV above $E_{\rm{F}}$ due to the low excitation energy.
 Curiously, the population of the TSS is pronounced at a specific energy
of $\sim$300~meV (red arrows).
 It becomes even more pronounced at 80~K [Fig.~\ref{fig1}~(c)], where also
a considerable population near the lower edge of bulk conduction band
(BCB) at 270~meV can be observed.
 The specific energy of the population enhancement in the TSS significantly
depends on $h\nu_1$: it clearly shifts to lower energy with
decreasing $h\nu_1$ even for small changes of $h\nu_1$ [Fig.~\ref{fig1}~(c)-(e)].
 The data also clearly shows that the population at $-k_{||}$ is
much more prominent compared to that at $+k_{||}$, which represents a strong
asymmetric population in $k$-space.
 The population enhancement and its strong asymmetry is not observed
for visible pump pulses ($h\nu_1$=2.58~eV) and the same UV probe pulses
[Fig.~\ref{fig1}~(f)].
 Thus, it turns out that theses findings are unique for the optical excitation
with MIR pulses.

\begin{figure}[b]
\begin{center}
\includegraphics[width=0.95\columnwidth]{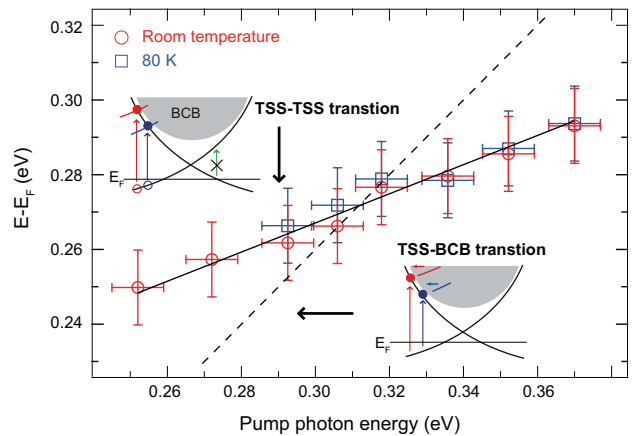}
\caption[]{(Color online)
  Energy position of the pronounced 2PPE signal created
  in the TSS as a function of the MIR photon energy acquired at 300~K
  (red circle) and 80~K (blue circle), respectively.
  The solid and the dashed line show the simulated energy dependence
  for TSS-TSS transitions and TSS-BCB transitions (see text for details).}
\label{fig2}
\end{center}
\end{figure}
In order to reveal the origin of the pronounced population at a specific energy,
we have investigated its dependence on $h\nu_1$ by systematic tuning of the MIR pulses.
 Figure~\ref{fig2} summarizes this dependence for 300~K (red symbols) and 80~K (blue symbols).
 Obviously, the energy position of the population enhancement is proportional
to the excitation energy for both temperatures.
 This is a clear indication that it is induced by a specific direct optical
coupling to the TSS.

In principle, three different optical excitation processes can result
in a population of the TSS within this energy range:
 Firstly, transitions from the bulk valence band (BVB) to the BCB.
 This process, however, can neither raise the population at a specific
energy in the TSS, nor should it show a clear $h\nu_1$ dependence, because both
bands can be coupled over a wide energy range.
Secondly, TSS-BCB (BVB-TSS) transitions for which the initial (intermediate) state
is the TSS and the intermediate (initial) state is a bulk state.
 Thirdly, resonant transitions between the occupied and unoccupied
part of the TSS across the DP (TSS-TSS transition).

\begin{figure*}[t!]
\begin{center}
\includegraphics[width=0.85\textwidth]{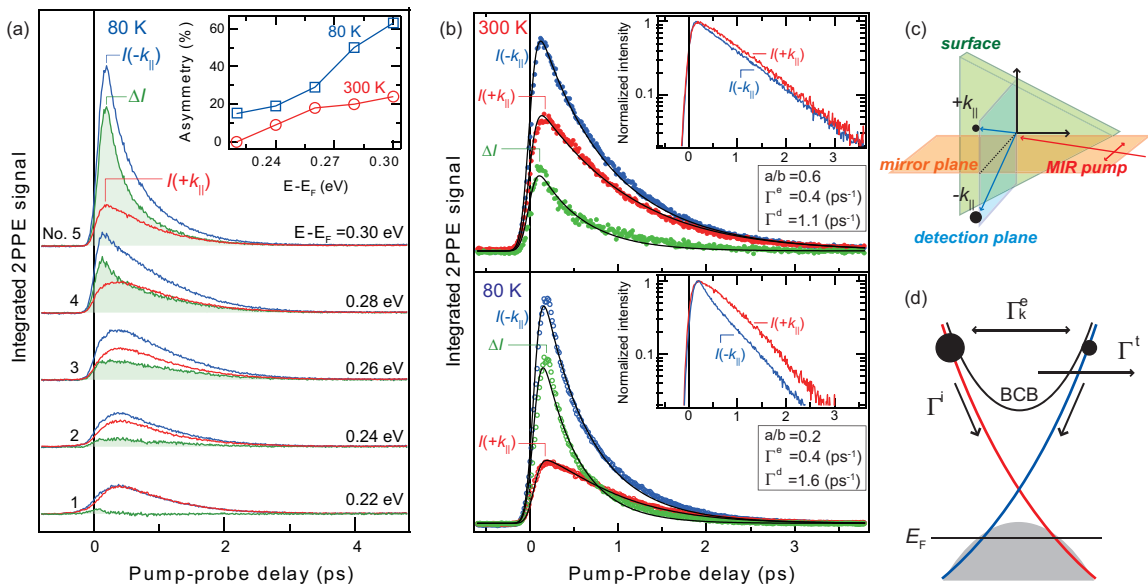}
\caption[]{(Color online)
  (a) Time-evolutions of the 2PPE signals $I_{+k}$ (red), $I_{-k}$ (blue)
  and their difference ${\Delta}I$=$I_{-k}-I_{+k}$ (green) for different
  energies at 80~K.
  $I_{\pm{k}}$ is respectively obtained by integrating the 2PPE signal within
  the windows shown in Fig.~\ref{fig1}~(c). The inset depicts the
  maximum of $A$=${\Delta}I/(I_{+k}+I_{-k})$ as a function of energy.
  (b) Representative line profiles from (a) at the direct excitation energy at
  300~K (top) and 80~K (bottom).
  Solid lines indicate the fitting results of the rate equation model [see the main text].
  The fitting parameters are indicated in the figure.
  The inset shows logarithmic plots of the normalized $I_{\pm{k}}$.
  (c) Experimental geometry where a mirror plane of the
      surface coincides with the plane of incidence (red square) perpendicular to
      the detection plane of photoelectrons (blue square).
       The triangle depicts the $C_{3v}$ surface symmetry.
   (d) Energy and momentum scheme for the decay of the asymmetric population
  due to elastic scattering within the TSS ($\Gamma^e_{k}$), inelastic scattering
  ($\Gamma^i$) and transport ($\Gamma^t$).
   }
\label{fig3}
\end{center}
\end{figure*}
It can be easily understood that TSS-BCB and BVB-TSS transitions, which are transitions
into (from) a continuum of states, should show a linear dependence on $h\nu_1$
with a slope of unity as depicted by the dashed line in Fig.~\ref{fig2}.
 Obviously, this dependence cannot describe our experimental data.
 The $h\nu_1$ dependence of TSS-TSS transitions, on the other hand,
is given by the dispersion of the TSS.
 The population enhancement for this process appears at those
intermediate state energies for which the energy difference between
the initial state in the occupied part of the TSS and the intermediate
state in the unoccupied part of the TSS just matches $h\nu_1$.
 For the simulation of this process, we have used data
on the dispersion of the occupied part of the TSS  in Sb$_{2}$Te$_{3}$
from Ref.~\cite{Pauly12prb} and for the unoccupied part from Ref.~\cite{Reimann14prb}.
 The almost linear dispersion of the TSS results in a simple linear
dependence on $h\nu_1$ as depicted by the solid line in Fig.~\ref{fig2}
which can excellently reproduce our experimental results.
 The simulation furthermore predicts that TSS-TSS
transitions are not possible for $h\nu_1$ below 0.27~eV at which
the initial state becomes unoccupied.
 This is also in very good agreement with our finding that
the population in the TSS strongly drops for $h\nu_1$ below 0.25~eV.
 It is therefore concluded that MIR pulses are able to drive a
direct optical transition from the lower into the upper part of the TSS.

One may wonder why such a transition between states of opposite
chiral spin textures across the DP are allowed,
since spin-flip excitations are forbidden in the dipole approximation.
 However, if spin-orbit coupling plays an important role,
the TSS cannot be fully spin polarized and this selection rule is softened.
 In fact, a spin polarization of below 80\% has been discussed for the TSS in different
3D TIs~\cite{Yazyev_spin_calc, Pan_PRL, Souma_spin, Miyamoto_spin}.
 In addition, a hybridization with bulk states~\cite{Seibel15prl}
possibly results in a further reduction of the spin polarization~\cite{Pauly12prb}.

We now turn to the asymmetry of the excited population which is only observed
for direct excitation with MIR pulses.
 Its time and energy evolution is shown in Fig.~\ref{fig3}~(a), where we have plotted the
transient 2PPE intensity at ${\pm}k$ as well as its difference for
different energies as depicted by the five integration windows in Fig.~\ref{fig1}~(c).
 For all energies, the 2PPE intensity shows a fast rise within the time resolution of $\approx 200$~fs
and a subsequent decay within a few picoseconds.
 This is in strong contrast to the delayed dynamics observed for 800-nm, or visible excitation
 \cite{Sobota12prl,Wang12prl,Crepaldi12prl,Hajlaoui12Nano,Sobota13prl,
  Niesner14prb,Reimann14prb},
and corroborates the direct excitation process of the TSS.
 The asymmetry $A=\Delta I/(I_{-k}+I_{+k})$ reaches up to 60\% at the direct excitation energy
and strongly drops for energies below.
 We observe such large asymmetry along $\bar{\Gamma}$-$\bar{\rm K}$
for $p$-polarized MIR pulses but almost no contrast between
opposite helicities of circular polarized light~\cite{SI}.
 This is surprising for our geometry where a mirror plane of the
surface coincides with the plane of incidence
[Fig.~\ref{fig3}~(c)].
 It indicates that the three-fold symmetry of the surface might
be broken.
 Possible reasons for such symmetry break can be oriented
step edges on the cleaved surface which superimpose a one-fold
symmetry, a distortion of the first quintuple layer with
respect to the underlying bulk, or a non-perfect azimuthal
orientation of the sample, which was, however, oriented within
better than 5$^\circ$~\cite{SI}.
 Independent of its actual origin, a k-space asymmetry of the
population in the intermediate state directly reflects a
photocurrent parallel to the surface \cite{Guedde07Science}.
 This is verified by the observation of a distinct dynamics  of
the 2PPE signal for opposite momenta at the direct excitation
energy [Fig.~\ref{fig3}~(b)], which is most clearly seen in the
logarithmic plot of the normalized intensities
 [inset of Fig.~\ref{fig3}~(b)].
 The
 decay at $-k$ is
initially faster as compared to the
 decay at $+k$ due to momentum scattering which progressively equalizes the asymmetry.
 On a longer timescale, both signals decay with a common time constant due to inelastic decay.
 In accordance with the intensity contrast, the difference of the decay dynamics is more pronounced at 80~K
as compared to 300~K [Fig.~\ref{fig3}~(b)].
 These findings unambiguously reveal that the novel direct optical excitation
generates a transient photocurrent in the TSS.

To analyze the dynamics of the photocurrent in more detail, we
use a rate-equation model which is depicted in
Fig.~\ref{fig3}~(d). It describes the populations $n_{+k}$ and
$n_{-k}$ at the direct excitation energy where no indirect
filling from higher-lying states can occur:
\begin{eqnarray*}
\frac{dn_{+}}{dt}&=&a{\delta}(t)-{\Gamma^{e}_{k}}{n_{+}}+{\Gamma^{e}_{k}}{n_{-}}-{\Gamma^{d}}{n_{+}}\\
\frac{dn_{-}}{dt}&=&b{\delta}(t)-{\Gamma^{e}_{k}}{n_{-}}+{\Gamma^{e}_{k}}{n_{+}}-{\Gamma^{d}}{n_{-}}
\label{eq:rate_carrier}
\end{eqnarray*}
Here, ${\delta}(t)$ is the temporal intensity profile of the
Gaussian shaped MIR laser pulse, and $a$ and $b$ indicate the
different excitation probabilities of electrons at $+k_{||}$
and $-k_{||}$, respectively. In our model, $n_{+k}$ and
$n_{-k}$ exchange electrons with an elastic scattering rate
($\Gamma^{e}_{k}$) which includes momentum and spin scattering.
Both mechanisms are closely related to each other due to the
spin structure of the Dirac cone but might only be disentangled
by a direct observation of the spin dynamics.
 Both populations mutually decay with an effective total population
decay rate ($\Gamma^{d}$).
 Beside the inelastic decay into lower-lying states ($\Gamma^{i}$),
$\Gamma^{d}$ includes interband scattering into the BCB with subsequent bulk
transport ($\Gamma^{t}$), because the direct excitation energy is close to the
BCB bottom~\cite{Reimann14prb}.
$\Gamma^{d}$ is thus defined as
$\Gamma^{d}=\Gamma^{i}+\Gamma^{t}$.

By assuming that $I_{\pm k}{\propto}n_{\pm k}$, the difference ${\Delta}{I}=I_{+k}-I_{-k}$
directly reflects the photocurrent in the TSS.
 With the two rate equations above, it is described by:
\begin{eqnarray*}
\frac{d{\Delta}I}{dt}&=&(b-a){\delta}(t)-({2\Gamma^{e}_{k}+\Gamma^{d}}){\Delta}I
\label{eq:rate_current}
\end{eqnarray*}
 The photocurrent thus decays exponentially with a time constant $\tau^{c}=1/({2\Gamma^{e}_{k}+\Gamma^{d}}$)
 which is governed by both elastic momentum scattering and total population decay.
 Best fits for this model of the experimental data $I_{\pm{k}}$ and ${\Delta}{I}$
are shown as solid lines in Fig.~\ref{fig3} (b).
 Clearly, even this simple model can reproduce the experimental data
for both temperatures very well.
 $\Gamma^{d}$ is for both temperatures much larger compared to $\Gamma^{e}_{k}$ through $\Gamma^{i}$,
because electron-hole pair creation in the incompletely filled VB
is an important inelastic decay channel for $p$-doped samples.
 $\tau^{c}$=0.42 (0.52)~ps at 80 (300)~K is therefore governed by the overall population decay of the TSS.
 The increase of $\Gamma^{d}$ at 80~K as compared to 300~K can be explained
by an enhancement of $\Gamma^{t}$,
which shows quantitatively good agreement with our recent work on
Sb$_{2}$Te$_{2}$Se~\cite{Reimann14prb}.
 In contrast, $\Gamma^{e}_{k}$ shows no significant change with temperature
although the Debye temperature of Sb$_2$Te$_3$ ($\theta_D=162$~K~\cite{Dyck02prb}) is
well between the two investigated temperatures.
 We thus conclude that phonon scattering plays only a minor role
for $\Gamma^{e}_{k}$ and surface imperfections like steps or
defects are most likely the main factors for elastic
scattering.
 Beyond defect scattering, interband scattering into and from
the BCB, where backscattering is allowed, might also be
possible due to the large wavefunction overlap of the TSS with
the BCB.
 Both scattering processes effectively increase $\Gamma^{e}_{k}$
which ultimately limits the lifetime of the photocurrent even if $\Gamma^{d}$ can be suppressed.
 Indeed, $\Gamma^{e}_{k}$ becomes the main factor for the transport
properties under static electric fields if the sample is close
to charge neutrality.
 In any case, the decay time $\tau^{e}_{k}=1/\Gamma^{e}_{k}=2.5$~ps is quite long,
if for example, compared to dephasing times of quantum beats
between image potential states on well prepared noble metal
surfaces~\cite{Hofer97sci,Reuss99prl,Marks11prb}.
 Such a slow randomization of momentum holds a general advantage
also for application under static conditions.
 If the main process of $\Gamma^{e}_{k}$ is in fact defect scattering,
high-quality thin film samples might further increase the mobility of
electrons in the TSS.

%
%
In conclusion, we have shown that MIR pulses generate a novel optical
coupling between the occupied and the unoccupied part of the TSS
across the DP, which permits an ultrafast direct excitation of the TSS.
 Our data unambiguously reveal that this novel direct excitation
generates an unbalanced transient electron population in
$k$-space, which directly mirrors a photoexcited electric
current in the TSS.
 Even if the polarization dependence of the current generation
is not fully understood yet, our discovery opens a pathway for
a coherent optical control of the TSS via an ultrafast optical
excitation with MIR pulses.

%
We thank H.~Bentmann and F.~Reinert for providing us Sb$_{2}$Te$_{3}$
samples and gratefully acknowledge funding by the Deutsche
Forschungsgemeinschaft through SPP1666 and GU495/2.
K.~K. acknowledges support from JSPS Postdoctral Fellowship for Research Abroad.
%

\bibliographystyle{prsty}

\end{document}